\documentclass[cameraready]{Interspeech}
\title{The False Resonance: A Critical Examination of Emotion Embedding Similarity for Speech Generation Evaluation}

\author[affiliation={1},orcid=0009-0001-0466-9364]{Yun-Shao}{Tsai}
\author[affiliation={1}, orcid=0009-0007-3994-6433]{Yi-Cheng}{Lin}
\author[affiliation={2}, orcid=0000-0003-2125-5689]{Huang-Cheng}{Chou}
\author[affiliation={3}]{Tzu-Wen}{Hsu}
\author[affiliation={4},orcid=0009-0001-6826-1747]{Yun-Man}{Hsu}
\author[affiliation={4},orcid=0009-0001-1889-2663]{Chun Wei}{Chen}
\author[affiliation={2},orcid=0000-0002-1052-6204]{Shrikanth}{Narayanan}
\author[affiliation={1,5}, orcid=0000-0002-9654-5747, correspondingauthor]{Hung-yi}{Lee}

\address{
    $^1$ Graduate Institute of Communication Engineering, National Taiwan University, Taiwan \\
    $^2$ Signal Analysis and Interpretation Laboratory (SAIL), University of Southern California, USA\\
    $^3$ Independent Researcher
    $^4$ National Taiwan University, Taiwan \\
    $^5$ Artificial Intelligence Center of Research Excellence, National Taiwan University, Taiwan
}

\email{r14942093@ntu.edu.tw, hungyilee@ntu.edu.tw}

\keywords{Emotion Similarity Measure, Text-to-Speech, Voice Conversion, Objective Evaluation, Emotion Embeddings}

\usepackage{comment}
\usepackage{amssymb}
\usepackage{microtype}
\usepackage{amsmath}
\usepackage{multirow}
\usepackage{cite}
\usepackage{pgfplots}
\usepgfplotslibrary{fillbetween}
\pgfplotsset{compat=1.17}
\usepgfplotslibrary{groupplots}
\newcommand{\sd}[1]{\pm \text{\tiny #1}} 
\newcommand{\m}{\scalebox{0.45}[1.0]{$-$}}
\newcommand{\dash}{\rule[0.5ex]{0.8em}{0.4pt}}
\begin{document}

\maketitle

\begin{abstract}
Objective metrics for emotional expressiveness are vital for speech generation, particularly in expressive synthesis and voice conversion requiring emotional prosody transfer.
To quantify this, the field widely relies on emotion similarity between reference and generated samples.
This approach computes cosine similarity of embeddings from encoders like \textit{emotion2vec}, assuming they capture affective cues despite linguistic and speaker variations.
We challenge this assumption through controlled adversarial tasks and human alignment tests.
Despite high classification accuracy, these latent spaces are unsuitable for zero-shot similarity evaluation.
Representational limitations cause linguistic and speaker interference to overshadow emotional features, degrading discriminative ability.
Consequently, the metric misaligns with human perception.
This acoustic vulnerability reveals it rewards acoustic mimicry over genuine emotional synthesis.
\end{abstract}

\begin{table*}[t]
\centering
\fontsize{8}{10}\selectfont 
\caption{\small \textbf{Categorical Emotion Evaluation}.
Triplet accuracy (\%) under controlled adversarial settings.
Values are Mean ± $\sd{SD}$ (dashes denote exclusion due to presence in pre-training corpora for zero-shot evaluation).
Performance frequently falls to or below random chance (50 \%), demonstrating a fundamental vulnerability to acoustic distractors.}
\vspace{-2mm}
\label{tab:categorical_results}
\begin{tabular}{lccccccc}
\toprule
\textbf{Dataset} & \textbf{emotion2vec (e2v)} & \textbf{e2v$^{+}$ seed} & \textbf{e2v$^{+}$ base} & \textbf{e2v$^{+}$ large} & \textbf{HuBERT} & \textbf{W2V 2.0} & \textbf{TERA} \\
\midrule
\multicolumn{8}{c}{\textbf{Unconstrained Sampling}} \\
\midrule
MSP-Imp.  & $55.28 \sd{0.96}$ & $63.02 \sd{1.97}$ & $58.08 \sd{1.90}$ & $62.08 \sd{2.40}$ & $54.70 \sd{1.18}$ & $53.08 \sd{2.91}$ & $56.40 \sd{1.05}$ \\
MSP-Pod.  & $51.94 \sd{1.22}$ & \dash & \dash & \dash & $55.92 \sd{2.13}$ & $53.74 \sd{1.72}$ & $54.92 \sd{3.04}$ \\
BIIC-Pod. & $51.08 \sd{1.16}$ & $55.64 \sd{1.43}$ & $52.56 \sd{1.45}$ & $53.62 \sd{0.73}$ & $54.06 \sd{1.50}$ & $53.02 \sd{1.52}$ & $53.58 \sd{1.07}$ \\
Dusha     & $52.74 \sd{2.62}$ & $64.78 \sd{0.83}$ & $60.12 \sd{1.56}$ & $65.66 \sd{1.42}$ & $52.26 \sd{0.96}$ & $52.22 \sd{2.01}$ & $53.54 \sd{2.57}$ \\
CREMA-D   & $55.66 \sd{0.74}$ & \dash & \dash & \dash & $64.22 \sd{1.14}$ & $64.68 \sd{2.26}$ & $67.30 \sd{1.15}$ \\
NNIME     & $53.14 \sd{3.08}$ & $55.54 \sd{1.58}$ & $53.40 \sd{1.04}$ & $55.22 \sd{1.65}$ & $53.98 \sd{1.59}$ & $53.12 \sd{1.93}$ & $54.06 \sd{1.74}$ \\
\midrule
\multicolumn{8}{c}{\textbf{Speaker-Linguistic Match}} \\
\midrule
MSP-Imp.  & $49.84 \sd{1.32}$ & $69.32 \sd{2.66}$ & $60.00 \sd{1.68}$ & $65.24 \sd{1.60}$ & $48.92 \sd{0.95}$ & $47.72 \sd{0.83}$ & $51.48 \sd{1.27}$ \\
CREMA-D   & $63.08 \sd{1.44}$ & \dash & \dash & \dash & $70.84 \sd{2.26}$ & $70.92 \sd{2.28}$ & $70.00 \sd{1.05}$ \\
\midrule
\multicolumn{8}{c}{\textbf{Speaker Distractor}} \\
\midrule
MSP-Imp.  & $37.86 \sd{1.27}$ & $57.82 \sd{1.16}$ & $51.92 \sd{1.54}$ & $55.16 \sd{1.26}$ & $31.88 \sd{1.73}$ & $31.22 \sd{1.37}$ & $29.86 \sd{1.59}$ \\
CREMA-D   & $43.94 \sd{1.38}$ & \dash & \dash & \dash & $43.32 \sd{1.87}$ & $49.74 \sd{0.94}$ & $50.98 \sd{1.78}$ \\
\midrule
\multicolumn{8}{c}{\textbf{Linguistic Distractor}} \\
\midrule
MSP-Imp.  & $20.14 \sd{1.25}$ & $53.48 \sd{1.71}$ & $53.08 \sd{2.04}$ & $50.10 \sd{1.57}$ & $35.32 \sd{1.59}$ & $44.80 \sd{1.96}$ & $47.14 \sd{1.71}$ \\
CREMA-D   & $3.38 \sd{0.27}$  & \dash & \dash & \dash & $28.46 \sd{0.92}$ & $50.62 \sd{1.23}$ & $58.84 \sd{0.79}$ \\
\bottomrule
\end{tabular}
\vspace{-2mm}
\end{table*}
\begin{table*}[t]
\fontsize{8}{10}\selectfont 
\centering
\caption{\small \textbf{Shift Discriminability (Accuracy)}. Performance comparison across four datasets.
Results are reported in Mean ± $\sd{SD}$ (\%) (dashes denote exclusion due to presence in pre-training corpora for zero-shot evaluation).
Accuracies remain near random chance (50 \%), exposing the metric's inability to track continuous emotion attributes.}
\vspace{-2mm}
\label{tab:shift_discriminability}
\setlength{\tabcolsep}{4pt}
\begin{tabular}{llccccccc}
\toprule
\textbf{Dataset} & \textbf{Dim.} & \textbf{emotion2vec (e2v)} & \textbf{e2v$^{+}$ seed} & \textbf{e2v$^{+}$ base} & \textbf{e2v$^{+}$ large} & \textbf{HuBERT} & \textbf{W2V 2.0} & \textbf{TERA} \\
\midrule

\multirow{2}{*}{MSP-Imp.} 
& Valence & $56.30 \sd{1.48}$ & $61.92 \sd{1.67}$ & $57.66 \sd{2.67}$ & $61.56 \sd{2.03}$ & $55.66 \sd{2.25}$ & $52.76 \sd{1.48}$ & $51.76 \sd{1.54}$ \\
& Arousal & $53.74 \sd{2.14}$ & $55.12 \sd{2.89}$ & $53.68 \sd{1.55}$ & $55.66 \sd{1.28}$ & $54.02 \sd{1.67}$ & $54.32 \sd{1.19}$ & $57.96 \sd{2.13}$ \\
\midrule

\multirow{2}{*}{MSP-Pod.} 
& Valence & $52.46 \sd{2.42}$ & \dash & \dash & \dash & $51.70 \sd{1.73}$ & $51.06 \sd{1.36}$ & $51.60 \sd{1.03}$ \\
& Arousal & $52.88 \sd{0.92}$ & \dash & \dash & \dash & $55.50 \sd{1.29}$ & $54.10 \sd{0.45}$ & $55.12 \sd{1.27}$ \\
\midrule

\multirow{2}{*}{BIIC-Pod.} 
& Valence & $52.20 \sd{0.62}$ & $54.16 \sd{1.38}$ & $51.48 \sd{1.13}$ & $53.22 \sd{1.44}$ & $54.20 \sd{2.31}$ & $51.90 \sd{1.87}$ & $52.58 \sd{1.59}$ \\
& Arousal & $52.64 \sd{2.92}$ & $57.04 \sd{1.51}$ & $52.90 \sd{1.12}$ & $56.54 \sd{0.50}$ & $53.02 \sd{1.30}$ & $52.68 \sd{1.81}$ & $52.32 \sd{1.01}$ \\
\midrule

\multirow{2}{*}{NNIME} 
& Valence & $51.82 \sd{0.50}$ & $53.94 \sd{1.22}$ & $54.16 \sd{2.65}$ & $54.40 \sd{1.94}$ & $52.88 \sd{1.74}$ & $51.84 \sd{1.06}$ & $54.40 \sd{1.59}$ \\
& Arousal & $54.98 \sd{1.26}$ & $56.30 \sd{2.39}$ & $53.92 \sd{1.97}$ & $57.32 \sd{0.46}$ & $57.76 \sd{0.87}$ & $58.82 \sd{0.81}$ & $61.56 \sd{0.75}$ \\

\bottomrule
\end{tabular}
\vspace{-3mm}
\end{table*}

\begin{table*}[t]
\centering
\caption{\small \textbf{Trend Monotonicity Evaluation.}
Spearman's rank correlation ($\rho$) across datasets.
Results are reported in Mean $\sd{SD}$ (dashes denote exclusion due to presence in pre-training corpora for zero-shot evaluation).
Near-zero correlations show the metric lacks a monotonic relationship with emotional magnitude.}
\vspace{-3mm}
\label{tab:trend_monotonicity}
\fontsize{8}{10}\selectfont 
\setlength{\tabcolsep}{4pt} 
\begin{tabular}{ll ccccccc}
\toprule
\textbf{Dataset} & \textbf{Dim.} & \textbf{emotion2vec (e2v)} & \textbf{e2v$^{+}$ seed} & \textbf{e2v$^{+}$ base} & \textbf{e2v$^{+}$ large} & \textbf{HuBERT} & \textbf{W2V 2.0} & \textbf{TERA} \\
\midrule

\multirow{2}{*}{MSP-Imp.} 
& Valence & $\m0.07 \sd{0.03}$ & $\m0.20 \sd{0.01}$ & $\m0.11 \sd{0.02}$ & $\m0.19 \sd{0.04}$ & $\m0.06 \sd{0.02}$ & $\m0.02 \sd{0.03}$ & $\m0.02 \sd{0.04}$ \\
& Arousal & $\m0.04 \sd{0.01}$ & $\m0.07 \sd{0.02}$ & $\m0.01 \sd{0.03}$ & $\m0.07 \sd{0.04}$ & $\m0.05 \sd{0.02}$ & $\m0.05 \sd{0.03}$ & $\m0.10 \sd{0.01}$ \\
\midrule

\multirow{2}{*}{MSP-Pod.} 
& Valence & $\m0.02 \sd{0.02}$ & \dash & \dash & \dash & $\m0.02 \sd{0.02}$ & $\m0.01 \sd{0.02}$ & $\m0.00 \sd{0.02}$ \\
& Arousal & $\m0.08 \sd{0.04}$ & \dash & \dash & \dash & $\m0.08 \sd{0.03}$ & $\m0.04 \sd{0.02}$ & $\m0.04 \sd{0.04}$ \\
\midrule

\multirow{2}{*}{BIIC-Pod.} 
& Valence & $\m0.03 \sd{0.02}$ & $\m0.08 \sd{0.02}$ & $\m0.04 \sd{0.02}$ & $\m0.06 \sd{0.01}$ & $\m0.04 \sd{0.04}$ & $\m0.03 \sd{0.03}$ & $\m0.03 \sd{0.05}$ \\
& Arousal & $\m0.03 \sd{0.03}$ & $\m0.11 \sd{0.04}$ & $\m0.06 \sd{0.04}$ & $\m0.10 \sd{0.03}$ & $\m0.03 \sd{0.02}$ & $\m0.03 \sd{0.02}$ & $\m0.03 \sd{0.02}$ \\
\midrule

\multirow{2}{*}{NNIME} 
& Valence & $\m0.06 \sd{0.05}$ & $\m0.05 \sd{0.05}$ & $\m0.05 \sd{0.01}$ & $\m0.10 \sd{0.02}$ & $\m0.05 \sd{0.02}$ & $\m0.03 \sd{0.02}$ & $\m0.05 \sd{0.02}$ \\
& Arousal & $\m0.06 \sd{0.04}$ & $\m0.04 \sd{0.02}$ & $\m0.03 \sd{0.02}$ & $\m0.09 \sd{0.02}$ & $\m0.06 \sd{0.04}$ & $\m0.08 \sd{0.03}$ & $\m0.12 \sd{0.02}$ \\

\bottomrule
\end{tabular}
\vspace{-6mm}
\end{table*}

\section{Introduction and Background}
\label{sec:intro}

As generative speech models rapidly evolve, generating emotionally expressive speech has become a key objective across tasks such as expressive text-to-speech (TTS) and emotional voice conversion (EVC) \cite{xie-etal-2025-towards, hsu2025breezyvoice}.
Consequently, evaluating how well generated audio captures the desired emotional style is essential.
While subjective mean opinion score testing remains the gold standard, its high cost and inherent rater variability have motivated the field to seek scalable, automated alternatives.

To bypass the limitations of subjective testing, evaluating synthesized speech increasingly relies on latent representations.
While \textit{WavLM-TDNN} \cite{9814838, zhang2024speechtokenizer} remains standard for speaker and vocal style similarity \cite{barrault2023seamless}, computing cosine similarity using embeddings from state-of-the-art speech emotion recognition (SER) models has rapidly become the de facto metric for quantifying emotion similarity (EMO-SIM).
The most widely adopted encoder is \textit{emotion2vec} \cite{ma-etal-2024-emotion2vec}.
This specific metric has seen widespread adoption across recent zero-shot TTS \cite{wang2025maskgct, zhou2025indextts2, wang2024emopro, 10965917, 10888737, 10832181, jeong2025tts, cho25b_interspeech, yang2025emovoice, michel2025libriquote, luo2025efficient, wang2025wordlevel, huang2024debatts, nie2025hd, shi2025emotion, borisov25_ssw}, EVC frameworks \cite{10844546, yoon2025maestro, du2025naturalvoices, dutta2025audio, wang2025beyond, pan25b_interspeech, gong2025perturbation}, and evaluation toolkits \cite{shi-etal-2025-versa}, marking a pervasive trend in the field \cite{lin24i_interspeech, lin24b_interspeech}.
However, high classification accuracy does not imply suitability for zero-shot similarity evaluation.
Most literature treats these embeddings as black boxes.
This practice implicitly presumes that spatial proximity reflects affective transfer, often without rigorously accounting for interference from speaker identity or linguistic content.

Building on these unverified assumptions, the widespread reliance on EMO-SIM poses a critical risk.
Automated metrics are routinely used for rapid iteration and model selection.
If an inaccurate metric rewards sound copying over real emotional expression, models will simply learn to duplicate speaker and linguistic details.
Consequently, relying on EMO-SIM risks guiding model development away from authentic affective speech before expensive MOS tests are ever conducted.

Inspired by recent perceptually relevant evaluation frameworks \cite{manku2025emergentttseval}, we argue a valid emotion similarity metric should satisfy three criteria:
(i) \textbf{Categorical Emotion Robustness}: accurately distinguishing discrete emotions despite acoustic interference,
(ii) \textbf{Dimensional Emotion Sensitivity}: maintaining high resolution and monotonic correlation with continuous dimensions (e.g., valence, arousal); 
(iii) \textbf{Human Perception Alignment}: reliably predicting subjective judgments.

Rather than evaluating synthesis systems, we subject the metric itself to rigorous attribute testing.
Our controlled experiments reveal significant limitations within these widely adopted encoders.
We summarize our contributions as follows:
\begin{itemize}
\item \textbf{Deficiencies in Robustness and Resolution}:
We assess EMO-SIM through adversarial sampling and continuous attribute tasks.
Even after calibrating the latent space to mitigate inherent representation crowding, our results demonstrate that the metric is severely compromised by acoustic distractors in categorical evaluation, and it fails to provide the resolution required to track fine-grained dimensional variations.
\item \textbf{Misalignment with Human Perception}: We evaluate the metric against human preferences using generated speech samples.
The results expose another significant gap, confirming that EMO-SIM fails to serve as a reliable objective proxy for human perception of emotional resemblance.
\item \textbf{Layer-wise Interference Tracking}: Through a layer-wise probing of \textit{emotion2vec}, we reveal that linguistic and speaker distractors severely disrupt the metric across all depths, while representations from deep layers actively degrade the alignment with human perceptual judgments. 
\end{itemize}

\section{Methodology}
\label{sec:method}

To rigorously evaluate EMO-SIM against the three criteria outlined in Section \ref{sec:intro}, we design a systematic pipeline.
We first calibrate the anisotropic latent space to prevent similarity distortion , followed by the specific experimental setups for our categorical, dimensional, and perceptual evaluations.

\subsection{Metric Calibration via Mean Centering}
\vspace{-1mm}
\label{ssec:metric_def}

While prior works compute cosine similarity directly on extracted embeddings, our preliminary analysis of 1,000 random audio pairs across each of our six experimental datasets reveals that the \textit{emotion2vec} latent space is highly anisotropic, consistent with findings in other self-supervised speech representations \cite{yang-etal-2022-self, wisniewski2025assessing}.
Uncalibrated similarity scores consistently range between 0.92 and 0.98, indicating that the embeddings occupy a narrow cone in the vector space.
This anisotropy severely compresses the metric's resolution.
To address this, we apply mean centering to shift the distribution to the origin, eliminating the common mean vector that dominates the representation \cite{mu2018allbutthetop}.
We compute the mean vector $\boldsymbol{\mu}$ over the specific dataset used for each evaluation run.
For indices $i, j \in \{1, 2, \dots, N\}$, let $\mathbf{e}_i$ and $\mathbf{e}_j$ be the utterance-level embeddings extracted from speech samples $x_i$ and $x_j$, respectively. 
The centered emotion similarity (EMO-SIM) is defined as:
\begin{equation}
\text{EMO-SIM}(x_i, x_j) = \frac{(\mathbf{e}_i - \boldsymbol{\mu}) \cdot (\mathbf{e}_j - \boldsymbol{\mu})}{\lVert \mathbf{e}_i - \boldsymbol{\mu} \rVert \lVert \mathbf{e}_j - \boldsymbol{\mu} \rVert}.
\end{equation}

\subsection{Categorical Emotion Evaluation}
\vspace{-1mm}
\label{ssec:categorical}

We assess \textit{EMO-SIM}'s discriminative power using a triplet task $\mathcal{T} = (x_{ref}, x_{pos}, x_{neg})$, where only $x_{pos}$ shares the reference $x_{ref}$'s emotion label.
The metric is accurate if $\text{EMO-SIM}(x_{ref}, x_{pos}) > \text{EMO-SIM}(x_{ref}, x_{neg})$.

To test robustness against acoustic distractors, we constrain speaker identity and linguistic content across 4 scenarios:
(i) \textbf{unconstrained sampling}: Samples are drawn randomly without constraints;
(ii) \textbf{speaker-linguistic match}: All triplet samples share the same speaker and linguistic content to strictly isolate affect;
(iii) \textbf{speaker distractor}: Linguistic content is fixed across the triplet, but only the negative sample matches the reference speaker (the positive sample features a different speaker);
(iv) \textbf{linguistic distractor}: Speaker identity is fixed across the triplet, but only the negative sample matches the reference linguistic content (the positive sample features different text).

\subsection{Dimensional Emotion Evaluation}
\vspace{-1mm}
\label{ssec:dimensional}

To evaluate sensitivity to dimensional affective attributes, we use valence ($s_{\text{val}}$) and arousal ($s_{\text{aro}}$) scores $s(x) \in [1, 5]$.

\noindent \textbf{Trend Monotonicity.} \quad
We measure the relationship between $\text{EMO-SIM}(x_i, x_j)$ and the score difference $|s(x_i) - s(x_j)|$ using Spearman's $\rho$.
For valence, we fix arousal for both samples; for arousal, we fix both valence and emotion category.
A valid metric should exhibit a negative correlation, indicating that similarity monotonically decreases as the score difference increases.

\noindent \textbf{Shift Discriminability.} \quad
We test the metric's capacity to discriminate clear score shifts using triplets $\mathcal{T} = (x_{ref}, x_{pos}, x_{neg})$ that satisfy the fixed-attribute constraints from the monotonicity analysis.
The positive sample exactly matches the reference score ($s(x_{pos}) = s(x_{ref})$ with strictly zero tolerance). This exact matching is feasible due to the annotation granularity of our selected datasets.
Meanwhile, the negative sample's score deviates by a margin of $|s(x_{neg}) - s(x_{ref})| \ge 1.0$.
For instance, an arousal triplet fixes valence (e.g., $s_{val} = 3.2$) and emotion (e.g., "Angry"). Given $s_{aro}(x_{ref}) = 4.8$, the positive sample exactly matches it ($s_{aro}(x_{pos}) = 4.8$), while the negative shifts distinctly (e.g., $s_{aro}(x_{neg}) = 2.5$). Accuracy is the proportion of triplets where the metric assigns higher similarity to $x_{pos}$.

\subsection{Human Perception Alignment Evaluation}
\vspace{-1mm}
\label{ssec:human_evaluation}

To assess alignment with human perception, we employ a pairwise preference task using triplets $\mathcal{T} = (x_{ref}, x_A, x_B)$, where $x_A$ and $x_B$ are synthetic utterances from different generative models.
Multiple researchers evaluate each triplet, with agreement quantified by Fleiss' $\kappa$\cite{Fleiss_1971}.
We define a valid dataset $\mathcal{D}_{valid}$ containing only triplets that reach a clear majority consensus.
For each triplet $\mathcal{T} \in \mathcal{D}_{valid}$, let $h \in \{x_A, x_B\}$ denote the human-preferred candidate, and $m = \operatorname*{argmax}_{x \in \{x_A, x_B\}} \text{EMO-SIM}(x_{ref}, x)$ denote the metric's choice.
Human alignment accuracy is the proportion of cases where $m = h$.

\section{Experimental Setup}
\label{sec:setup}

\subsection{Representation Extractors}
\vspace{-1mm}
\label{ssec:representation}

We evaluate the base \textit{emotion2vec} alongside its fine-tuned \textit{emotion2vec}$^{+}$ variants (seed, base, and large) \cite{ma-etal-2024-emotion2vec}.
We also include SSL models: \textit{HuBERT} \cite{9585401}, \textit{Wav2vec 2.0} \cite{NEURIPS2020_92d1e1eb}, and \textit{TERA} \cite{9478264}.
As in \textit{VERSA} \cite{shi-etal-2025-versa}, we extract frame-level representations from the last hidden layer, applying temporal mean pooling and mean centering calibration in Section \ref{ssec:metric_def}\footnote{Results using uncalibrated embeddings are consistently inferior or comparable to those reported.}.

\begin{table}[!t]
\caption{\small Utterance counts and distribution across selected databases. The `Selected' (Sel.) column represents data used after zero-shot filtering.}
\vspace{-3mm}
\label{tab:data_stats}
\centering
\fontsize{8}{10}\selectfont 
\setlength{\tabcolsep}{3.0pt} 
\begin{tabular}{lrrrrrr}
\toprule
\textbf{Database} & \textbf{Neu.} & \textbf{Hap.} & \textbf{Sad} & \textbf{Ang.} & \textbf{Sel.} & \textbf{Ori.} \\
\midrule
CREMA-D     & 3,897 & 353 & 370 & 986 & 5,606  & 7,442 \\
MSP-Imp.    & 3372 & 2603 & 873 & 767 & 7615  & 8,438 \\
MSP-Pod.  & 48,378 & 36,692 & 11,486 & 17,515 & 114,071 & 149,307 \\
BIIC-Pod. & 24,099 & 24,703 & 6,489 & 6,300 & 61,591 & 70,000 \\
Dusha     & 9,880 & 1,422 & 2,477 & 1,625 & 15,404 & 303,963 \\
NNIME      & 1,883 & 480 & 286 & 555 & 3,204  & 5,596 \\
\bottomrule
\end{tabular}
\vspace{-6mm}
\end{table}
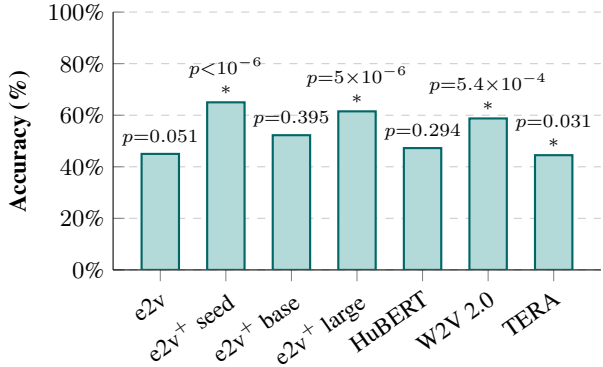
\begin{figure}[t]
\centering
\begin{tikzpicture}
\begin{axis}[
    ybar,
    width=\linewidth,         
    height=5cm,             
    bar width=14pt,           
    bar shift=0pt,
    ylabel={\textbf{Accuracy (\%)}},
    ymin=0, ymax=100,         
    ytick={0, 20, 40, 60, 80, 100}, 
    yticklabel={\pgfmathprintnumber{\tick}\%},
    symbolic x coords={e2v, e2v$^+$ seed, e2v$^+$ base, e2v$^+$ large, HuBERT, W2V 2.0, TERA},
    xtick=data,
    x tick label style={rotate=35, anchor=north east, font=\scriptsize, inner sep=2pt},
    ymajorgrids=true,         
    grid style={dashed, gray!40},
    axis x line*=bottom,
    axis y line*=left,
    enlarge x limits=0.12
]

\addplot[fill=teal!30, draw=teal!80!black, thick, forget plot] coordinates {
    (e2v, 45.00)
    (e2v$^+$ seed, 65.00)
    (e2v$^+$ base, 52.25)
    (e2v$^+$ large, 61.50)
    (HuBERT, 47.25)
    (W2V 2.0, 58.75)
    (TERA, 44.50)
};

\node[anchor=south, font=\scriptsize, align=center] at (axis cs:e2v, 45.00) {$p{=}0.051$};
\node[anchor=south, font=\scriptsize, align=center] at (axis cs:e2v$^+$ seed, 65.00) {$p{<}10^{-6}$\\$*$};
\node[anchor=south, font=\scriptsize, align=center] at (axis cs:e2v$^+$ base, 52.25) {$p{=}0.395$};
\node[anchor=south, font=\scriptsize, align=center] at (axis cs:e2v$^+$ large, 61.50) {$p{=}5{\times}10^{-6}$\\$*$};
\node[anchor=south, font=\scriptsize, align=center] at (axis cs:HuBERT, 47.25) {$p{=}0.294$};
\node[anchor=south, font=\scriptsize, align=center] at (axis cs:W2V 2.0, 58.75) {$p{=}5.4{\times}10^{-4}$\\$*$};
\node[anchor=south, font=\scriptsize, align=center] at (axis cs:TERA, 44.50) {$p{=}0.031$\\$*$};

\end{axis}
\end{tikzpicture}
\vspace{-3mm}
\caption{\small \textbf{Human Perception Alignment.} Accuracy (\%) of various models evaluated on human perception annotations. $*$ indicates statistical significance ($p < 0.05$) compared to the 50\% random baseline via binomial test.}
\label{fig:human_perception}
\vspace{-6mm}
\end{figure}

\subsection{Resource and Selection}
\vspace{-1mm}
\label{ssec:resource}

\noindent \textbf{Dataset Preparation.} \quad
We evaluate EMO-SIM across six diverse speech corpora spanning three languages: English (\textit{CREMA-D} \cite{6849440}, \textit{MSP-Improv} \cite{7374697}, \textit{MSP-Podcast} \cite{8003425,busso2025msppodcastcorpus}), Chinese (\textit{BIIC-Podcast} \cite{10388175}, \textit{NNIME} \cite{8273615}), and Russian (\textit{Dusha}, crowd test set only \cite{kondratenko2022large}).
To establish a standardized affective space across these varied sources, we filter the annotations to the intersection of four core categorical emotions: neutral, happy, sad, and angry.
This process involves mapping dataset-specific labels, such as ``positive'' in \textit{Dusha} and ``joy/excited'' in \textit{NNIME}, into the unified happy category.
Furthermore, to enforce a zero-shot evaluation setting, we exclude any datasets present in the pre-training corpora of the evaluated encoders.
The resulting data distributions are detailed in Table \ref{tab:data_stats}.
Finally, dataset assignment for each evaluation task is strictly governed by the availability of requisite annotations.
A corpus is utilized only if it possesses the requisite labels for a specific evaluation.

\noindent \textbf{Selection Strategy.} \quad
To ensure statistical robustness, we perform five independent sampling runs per dataset.
Each run constructs 1,000 unique evaluation instances (triplets or pairs), though this volume is adjusted to 500 for the speaker-linguistic match task due to the scarcity of exact parallel utterances.
For categorical and arousal evaluations, we enforce a balanced distribution of 25\% per emotion category.
Conversely, we relax this constraint for valence evaluation to avoid distorting natural affective correlations, sampling instead based on Section \ref{ssec:dimensional}.

\subsection{Human Perception Alignment Setup}
\label{ssec:human_study}
We initially constructed a large pool of preference triplets by drawing reference utterances ($x_{ref}$) from \textit{CREMA-D} (representing acted emotions) and \textit{MSP-Improv} (representing real-world spontaneous emotion).
For each triplet ($x_{ref}, x_A, x_B$), candidates were synthesized by models selected from \textit{CosyVoice} \cite{du2024cosyvoice}, \textit{SparkTTS} \cite{wang2025spark}, \textit{F5/E2 TTS} \cite{chen-etal-2025-f5, 10832320}, \textit{Qwen3-TTS} \cite{hu2026qwen3}, \textit{MaskGCT} \cite{wang2025maskgct}, \textit{Diff-HierVC} \cite{choi23d_interspeech}, and \textit{FreeVC} \cite{10095191}.
To isolate affective transfer, $x_A$ and $x_B$ both cloned the voice of $x_{ref}$ and shared identical linguistic content.
Five researchers evaluated each triplet by selecting from three options: Candidate A, Candidate B, or ``hard to distinguish''.
They were instructed to identify the candidate best matching the reference's emotion, regardless of quality.
The inter-rater agreement across the initial pool reached Fleiss' $\kappa = 0.7349$.
From the subset of triplets with strong consensus ($\ge 4/5$ agreement), we retained exactly 400 triplets (200 per source dataset) to ensure a balanced evaluation across both acted and spontaneous scenarios.

\section{Results and Analyses}
\label{sec:results}
\vspace{-1mm}
\subsection{Categorical Emotion Robustness}
\vspace{-1mm}
\label{ssec:categorical_analysis}

Table \ref{tab:categorical_results} reveals EMO-SIM's failure to capture genuine emotion.
Even in the ideal \textit{speaker-linguistic match} scenario, \textit{emotion2vec} and \textit{emotion2vec$^+$} variants barely reach 60-70\% accuracy.
This weak inherent affective representation severely degrades under adversarial acoustic variations.
With a \textit{linguistic distractor}, \textit{emotion2vec} accuracy plummets to 3.38\% on \textit{CREMA-D}.
Crucially, under both \textit{linguistic} and \textit{speaker distractors}, \textit{emotion2vec} and general SSL models frequently drop below chance.
This sub-random performance means the metric actively penalizes the correct emotion if the acoustic features differ, heavily favoring acoustic mimicry.
While \textit{emotion2vec$^+$} variants show slight resilience, they still degrade toward random guessing, solidifying EMO-SIM as an invalid proxy for zero-shot evaluation.

\subsection{Dimensional Emotion Sensitivity}
\vspace{-1mm}
\label{ssec:dimensional_analysis}

Tables \ref{tab:shift_discriminability} and \ref{tab:trend_monotonicity} reveal that EMO-SIM fails to capture continuous affective dimensions.
In \textit{shift discriminability}, \textit{emotion2vec} and \textit{emotion2vec$^+$} barely exceed random chance when distinguishing relative emotional shifts.
More severely, \textit{trend monotonicity} fails entirely: Spearman's rank correlation ($\rho$) remains near zero for both valence and arousal across all datasets.
This highlights an inability to map cosine similarity to continuous emotional magnitudes.
Ultimately, the metric is insensitive to fine-grained affective changes, confirming that current latent spaces lack the necessary resolution for dimensional emotion evaluation.

\subsection{Divergence from Human Perception}
\vspace{-1mm}
A valid objective metric should align with human subjective judgments.
Figure \ref{fig:human_perception} reveals that EMO-SIM falls significantly short of this standard.
Even the fine-tuned \textit{emotion2vec$^+$} variants achieve alignment accuracies between 52.25\% and 65.00\%.
While some variants marginally exceed chance, this performance remains inadequate for a proxy of human evaluation.
These results suggest the embedding space fails to capture nuanced affective prosody, likely relying on superficial acoustic similarities over genuine emotional equivalence.
Consequently, EMO-SIM is inherently unreliable as an automated perceptual metric for speech generation.

\begin{figure}[t]
\centering
\begin{tikzpicture}
\begin{axis}[
    width=\linewidth,
    height=5cm,
    xlabel={\textbf{Transformer Layers}},
    ylabel={\textbf{Accuracy (\%)}},
    ymin=0, ymax=100,       
    xmin=0, xmax=7,         
    xtick={0,1,2,3,4,5,6,7},
    xticklabels={L0, L1, L2, L3, L4, L5, L6, L7},
    ytick={0,20,40,60,80,100}, 
    yticklabel={\pgfmathprintnumber{\tick}\%}, 
    grid=major,             
    grid style={dashed, gray!30},
    legend style={
                at={(0.5, 1.05)}, 
                anchor=south,
                legend columns=2,  
                font=\scriptsize,
                cells={anchor=west},
                column sep=0.2ex,  
                draw=none,
                fill=white,
                fill opacity=0.8,
                text opacity=1
            },
    tick label style={font=\footnotesize},
    label style={font=\small}
]

\addplot[dashed, thick, gray] coordinates {(0,50) (7,50)};
\addlegendentry{Random Guess}

\addplot[name path=slm_upper, draw=none, forget plot] coordinates {(0, 52.32) (1, 52.61) (2, 51.85) (3, 51.88) (4, 51.84) (5, 51.83) (6, 52.14) (7, 51.16)};
\addplot[name path=slm_lower, draw=none, forget plot] coordinates {(0, 50.08) (1, 50.19) (2, 49.83) (3, 49.32) (4, 49.04) (5, 49.13) (6, 48.82) (7, 48.52)};
\addplot[purple, opacity=0.3, forget plot] fill between[of=slm_upper and slm_lower];
\addplot[color=purple, mark=star, thick, mark size=2.5pt] coordinates {(0, 51.20) (1, 51.40) (2, 50.84) (3, 50.60) (4, 50.44) (5, 50.48) (6, 50.48) (7, 49.84)};
\addlegendentry{Speaker-Ling. Match}

\addplot[name path=uncon_upper, draw=none, forget plot] coordinates {(0, 57.24) (1, 57.41) (2, 56.37) (3, 55.81) (4, 56.89) (5, 56.59) (6, 56.15) (7, 56.24)};
\addplot[name path=uncon_lower, draw=none, forget plot] coordinates {(0, 54.32) (1, 54.83) (2, 54.79) (3, 54.99) (4, 55.27) (5, 55.13) (6, 54.81) (7, 54.32)};
\addplot[blue, opacity=0.3, forget plot] fill between[of=uncon_upper and uncon_lower];
\addplot[color=blue, mark=*, thick, mark size=1.5pt] coordinates {(0, 55.78) (1, 56.12) (2, 55.58) (3, 55.40) (4, 56.08) (5, 55.86) (6, 55.48) (7, 55.28)};
\addlegendentry{Unconstrained}

\addplot[name path=spk_upper, draw=none, forget plot] coordinates {(0, 35.68) (1, 36.43) (2, 36.94) (3, 37.00) (4, 37.08) (5, 37.42) (6, 38.68) (7, 39.14)};
\addplot[name path=spk_lower, draw=none, forget plot] coordinates {(0, 32.88) (1, 34.61) (2, 33.34) (3, 33.72) (4, 34.40) (5, 34.86) (6, 35.32) (7, 36.58)};
\addplot[orange, opacity=0.3, forget plot] fill between[of=spk_upper and spk_lower];
\addplot[color=orange, mark=triangle*, thick, mark size=2pt] coordinates {(0, 34.28) (1, 35.52) (2, 35.14) (3, 35.36) (4, 35.74) (5, 36.14) (6, 37.00) (7, 37.86)};
\addlegendentry{Spk. Distractor}

\addplot[name path=lin_upper, draw=none, forget plot] coordinates {(0, 32.60) (1, 26.99) (2, 19.32) (3, 20.33) (4, 22.78) (5, 18.25) (6, 15.19) (7, 21.38)};
\addplot[name path=lin_lower, draw=none, forget plot] coordinates {(0, 30.68) (1, 23.57) (2, 16.88) (3, 18.07) (4, 20.50) (5, 15.95) (6, 13.17) (7, 18.90)};
\addplot[red, opacity=0.3, forget plot] fill between[of=lin_upper and lin_lower];
\addplot[color=red, mark=diamond*, thick, mark size=2pt] coordinates {(0, 31.64) (1, 25.28) (2, 18.10) (3, 19.20) (4, 21.64) (5, 17.10) (6, 14.18) (7, 20.14)};
\addlegendentry{Ling. Distractor}

\addplot[color=green!60!black, mark=square*, thick, mark size=1.5pt] coordinates {(0, 58.00) (1, 57.50) (2, 57.50) (3, 55.00) (4, 54.25) (5, 54.50) (6, 54.00) (7, 45.00)};
\addlegendentry{Human Perception}

\end{axis}
\end{tikzpicture}
\vspace{-7mm}
\caption{\small \textbf{Categorical Shift Tracking.} The trajectory of accuracy across transformer layers for the base \textit{emotion2vec} on MSP-Improv. Shaded regions denote standard deviation.}
\label{fig:layer_wise_triplet}
\vspace{-6mm}
\end{figure}
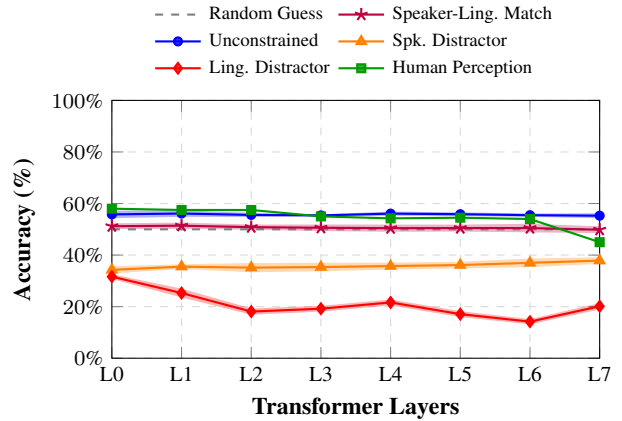

\subsection{Layer-Wise Analysis}
\vspace{-1mm}
\label{ssec:layer_analysis}

We probe the eight core Transformer blocks ($L0$–$L7$) of \textit{emotion2vec}, excluding the initial modality-specific adapters, to focus on the primary layers responsible for high-level affective representation.
Figure \ref{fig:layer_wise_triplet} reveals a striking structural deficiency: even without distractors, accuracy barely clears random chance across all depths.
Crucially, introducing acoustic distractors causes performance to plummet well below 50\% (e.g., dropping below 20\% under the \textit{Linguistic Distractor}), demonstrating that the latent space actively penalizes correct emotional pairs if their acoustic identities differ.
This vulnerability extends to human perception; rather than retaining affective features, deeper layers actively suppress them, causing alignment with human judgments to degrade from 58.0\% at L0 to a sub-random 45.0\% by L7.
This failure is further evidenced by dimensional sensitivity analysis, where Spearman’s rank correlation ($\rho$) for both arousal and valence hovers near zero at every depth.
Consequently, this persistent interference and diminishing perceptual alignment confirm that the core representations are inherently unsuited for zero-shot emotion evaluation.

\section{Discussion and Suggestion}
\label{sec:discuss}
\vspace{-1mm}
Our findings demonstrate that high SER accuracy does not inherently translate to a perceptually meaningful latent space.
We hypothesize \textit{emotion2vec} directly inherits acoustic representations from its foundational model, \textit{data2vec} \cite{baevski2022data2vec, waheed2024speech}.
While supervised SER utilizes linear classification layers as filters to suppress these non-affective acoustic properties, zero-shot EMO-SIM relies exclusively on cosine similarity, leaving spatial distances overwhelmingly dominated by non-affective features.

This \textit{false resonance} inadvertently rewards acoustic mimicry over genuine emotional transfer.
Because automated metrics drive rapid iteration and model selection, relying on such a misleading metric risks optimizing for acoustic cloning prior to subjective MOS testing.
To establish valid metrics, future work would benefit from exploring representation calibration.
For instance, introducing contrastive learning objectives during encoder training might help suppress acoustic variations and encourage distinct emotional clusters, potentially reshaping the embedding space to better reflect genuine affective prosody.

\section{Conclusion}
\label{sec:concl}
\vspace{-1mm}
EMO-SIM has become the de facto objective metric for zero-shot expressive speech evaluation.
However, this paper demonstrates that it is fundamentally unreliable.
Our empirical analyses reveal two critical flaws: current emotion embeddings are structurally biased by acoustic distractors, and their similarity scores severely misalign with human perception.
To address this \textit{false resonance}, future evaluation frameworks need to be designed to genuinely quantify emotional synthesis rather than penalizing it.

\section{Acknowledgments}
This work was supported by the Ministry of Education (MOE) Taiwan through NTU AI-CoRE; NSTC Taiwan (Grant 114-2917-I-564-030 to Huang-Cheng Chou); US NSF (IIS 2311676); and ODNI IARPA ARTS (Contract D2023-2308110001). The authors thank Sudarsana Reddy Kadiri for valuable feedback.


\section{Generative AI Use Disclosure}

We employed generative AI solely to improve the writing quality of this manuscript, without using it to generate any core content.

\bibliographystyle{IEEEtran}
\bibliography{mybib}

\end{document}